\documentclass[twocolumn,nofootinbib]{revtex4-1}
\usepackage[dvipdfmx]{graphicx}
\usepackage{dcolumn}
\usepackage{bm}
\usepackage{amsmath,amssymb,amsfonts}
\usepackage{latexsym}
\usepackage{bbm}
\usepackage[dvipdfmx]{color} 
\begin{document}


\title{Electroweak Baryogenesis with Lepton Flavor Violation}

\author{Cheng-Wei Chiang$^{1,2,3,4}$}%
\email{chengwei@phys.ntu.edu.tw}

\author{Kaori Fuyuto$^5$}
\email{ss6206@cc.saga-u.ac.jp}
\author{Eibun Senaha$^{1,2}$}
\email{senaha@ncu.edu.tw}
\affiliation{$^1$Department of Physics, National Taiwan University, Taipei 10617, Taiwan}
\affiliation{$^2$Department of Physics and Center for Mathematics and Theoretical Physics,
National Central University, Taoyuan 32001, Taiwan}
\affiliation{$^3$Institute of Physics, Academia Sinica, Taipei 11529, Taiwan}
\affiliation{$^4$Physics Division, National Center for Theoretical Sciences, Hsinchu 30013, Taiwan}
\affiliation{$^5$Department of Physics, Saga University, Saga 840-8502, Japan}
\bigskip

\date{\today}

\begin{abstract}
We investigate the feasibility of electroweak baryogenesis 
in a two-Higgs doublet model with lepton flavor violation. 
By scrutinizing the heavy Higgs boson mass spectrum, 
regions satisfying both strong first-order electroweak phase transition 
and the muon $g-2$ anomaly are identified. 
We also estimate the baryon number density by exploiting extra Yukawa couplings 
in the $\mu$-$\tau$ sector.
It is found that a CP-violating source term can be enhanced 
by the $\mu$-$\tau$ flavor-violating coupling together with the extra $\tau$ coupling.
With $\mathcal{O}(1)$ Yukawa couplings and CP-violating phases,
the observed baryon number density is marginally produced
under a generous assumption for the bubble wall profile. 
\end{abstract}

\maketitle

\section{Introduction}\label{sec:Intro}

Baryon asymmetry of the Universe (BAU) is an observational fact~\cite{Ade:2013zuv} whose origin still remains open and calls for physics beyond the standard model (SM).
Two important ingredients, namely, CP violation and departure from thermal equilibrium, are known to be insufficient in the SM~\cite{KMphase,lattice}.
A great number of baryogenesis mechanisms have been proposed so far, 
and the relevant energy scales are highly model-dependent. 
From the view point of testability, electroweak baryogenesis (EWBG)~\cite{ewbg,Morrissey:2012db}
is the most attractive scenario since it can be probed by current and foreseeable future experiments. 
It is thus interesting and important to scrutinize its feasible parameter space.

One of the simplest extensions for successful EWBG is to add another Higgs doublet to 
the SM Higgs sector, rendering
the so-called two-Higgs doublet model (2HDM) (for a review, see Ref.~\cite{Branco:2011iw}
and references therein).
In this model, both Higgs doublets can couple to quarks and leptons concurrently, 
giving rise to Higgs-mediated flavor-changing processes at tree level.
Importantly, flavor-changing neutral Higgs (FCNH) couplings are in general complex and may yield CP violation relevant to baryogenesis.
Earlier studies on EWBG with FCNH interactions can be found in Refs.~\cite{Tulin:2011wi,Cline:2011mm,Liu:2011jh}.
Moreover, it is possible to have a strong first-order electroweak phase transition (EWPT)
owing to the contributions of the additional Higgs doublet~\cite{EWPT_2HDM_1,EWPT_2HDM_2,Kanemura:2004ch,Fuyuto:2015vna}.

Recently, the CMS Collaboration reported an excess in a lepton flavor-violating (LFV) Higgs decay, ${\rm Br}(h\to \mu \tau) = (0.84^{+0.39}_{-0.37}) \%$, showing a 2.4$\sigma$ deviation from the SM~\cite{Khachatryan:2015kon}.\footnote{The excess would be less significant if the latest LHC Run-II data are taken into account~\cite{Cepeda}. Statistically, it is not yet sufficient to claim that the excess is completely gone.}  The ATLAS Collaboration, on the other hand, quotes an upper bound ${\rm Br}(h\to \mu \tau) < 1.43 \%$~\cite{Aad:2016blu}.
The authors of Refs.~\cite{mutau_2HDM_OST,Tobe:2016qhz}, including one of the present authors, pointed out that 
the 2HDM with $\mu$-$\tau$ flavor violation granted a parameter space to explain not only the $h\to \mu\tau$ excess
but also the long-standing anomaly in muon $g-2$ without having conflicts with other 
LFV constraints, such as $\tau\to\mu\gamma$, $\tau\to \mu\nu\bar{\nu}$, etc.
Since such FCNH interactions are generically accompanied by additional CP-violating sources, 
it is an interesting question whether the new CP-violating phases play a crucial role in
generating the BAU.

In this paper, we examine the possibility of EWBG in the 2HDM with $\mu$-$\tau$ flavor violation.
Our analysis is twofold: (1) baryon number generation via the new CP violation and 
(2) baryon number preservation by a strong first-order EWPT.
For the former, we evaluate the CP-violating source based on the method of closed-time path formalism
with vacuum expectation value (VEV) insertion, and then solve the diffusion equations
to estimate the BAU.
For the latter, we employ the one-loop finite temperature effective potential with a thermal resummation 
to determine the order of EWPT.
To determine the strength of first-order EWPT, we explicitly evaluate the energy of sphaleron configuration.

The paper is organized as follows.
In Sec.~\ref{sec:model}, we give basic ingredients of the 2HDM 
and show a relationship between FCNH couplings in the symmetric phase and those in the
broken phase.  In Sec.~\ref{sec:LFV}, we review how $h\to \mu\tau$ and $(g-2)_\mu$
can be simultaneously explained by the $\mu$-$\tau$ flavor violation,
as found in Refs.~\cite{mutau_2HDM_OST,Tobe:2016qhz}.
The baryon number preservation condition is given in Sec.~\ref{sec:BP}. 
We identify the regions where both strong first-order EWPT 
and $(g-2)_\mu$ anomaly can be accounted for. 
In Sec.~\ref{sec:BAU}, we present the BAU calculation and its numerical analysis.
Our conclusion is given in Sec.~\ref{sec:conclusion}.

\section{Model}\label{sec:model}

The 2HDM is an extension of the SM by adding one additional Higgs doublet. 
Such a two-Higgs doublets structure is motivated by some UV theories 
such as supersymmetric theories.
The most general Higgs potential at the renormalizable level is
\begin{align}
\lefteqn{V_0(\Phi_{1}, \Phi_{2}) }\nonumber\\
&=m_{1}^{2}\Phi_{1}^\dagger\Phi_1
	+m_{2}^{2}\Phi_{2}^\dagger\Phi_2
	-(m_{3}^{2}\Phi_{1}^{\dagger}\Phi_{2}+\mbox{h.c.}) \nonumber \\
&\quad	
	+\frac{\lambda_{1}}{2}(\Phi_1^\dagger\Phi_1)^2
	+\frac{\lambda_{2}}{2}(\Phi_2^\dagger\Phi_2)^2 
	+\lambda_{3}(\Phi_1^\dagger\Phi_1)(\Phi_2^\dagger\Phi_2) \nonumber\\
&\quad+\lambda_{4}(\Phi_1^\dagger\Phi_2)(\Phi_2^\dagger\Phi_1)
+\bigg[
	\frac{\lambda_{5}}{2}(\Phi_{1}^{\dagger}\Phi_{2})^{2} \nonumber\\
&\quad
	+\Big\{\lambda_6(\Phi_1^\dagger\Phi_1)
	+\lambda_7(\Phi_2^\dagger\Phi_2)
	\Big\}(\Phi_1^\dagger\Phi_2)
	+{\rm h.c.}
\bigg] ~.
\end{align}
Hermiticity of $V_0$ requires that $m_{1}^{2}$, $m_{2}^{2}$, 
$\lambda_{1}$, $\lambda_{2}$, $\lambda_{3}$ and $\lambda_{4}$ are all real 
while $m_{3}^{2}$, $\lambda_{5}$, $\lambda_{6}$ and $\lambda_{7}$ are generally complex, with one of them being possibly made real by a field redefinition of either Higgs doublet.
We parametrize $\Phi_i$ ($i=1,2$) as
\begin{align}
\Phi_{i}(x)=e^{i\theta_i}
\left(
\begin{array}{c}
\phi_{i}^{+}(x)\\
\frac{1}{\sqrt{2}}\Big(v_{i}+h_{i}(x)+ia_{i}(x)\Big)
\end{array}
\right) ~,
\end{align}
where $v_ie^{i\theta_i}$ are the VEV's.
For simplicity, we assume that CP is not violated by the Higgs potential and the VEV's,
leading to $\theta_{i}=0$.
Here we express $v_{1,2}$ in terms of polar coordinates, $v_1=v\cos\beta$ and $v_2=v\sin\beta$ with $v\simeq 246$ GeV. In the following, we will use the shorthand notation 
$s_\beta=\sin\beta$, $c_\beta=\cos\beta$ and $t_\beta=\tan\beta$.

In phenomenological discussions, it is useful to use the Higgs (Georgi) basis~\cite{HGbasis} in which
only one Higgs doublet develops the VEV and the Nambu-Goldstone bosons ($G^0,~G^\pm$) are decoupled from the physical states: 
\begin{align}
\begin{split}
\Phi'_1 &= c_\beta\Phi_1+s_\beta\Phi_2=
\left(
\begin{array}{c}
G^+ \\
\frac{1}{\sqrt{2}}\Big(v+h_1'+iG^0\Big)
\end{array}
\right) ~, \\
\Phi'_2 &= -s_\beta\Phi_1+c_\beta\Phi_2=
\left(
\begin{array}{c}
H^+ \\
\frac{1}{\sqrt{2}}\Big(h_2'+iA\Big)
\end{array}
\right) ~,
\end{split}
\end{align}
where $h_1'=c_{\beta-\alpha} H+s_{\beta-\alpha}h$ and $h_2'=-s_{\beta-\alpha}H+c_{\beta-\alpha}h$
with $\alpha$ being a mixing angle between the two CP-even Higgs fields ($h_{1,2}$).
In this paper, $h$ is the 125 GeV Higgs boson and assumed to be the lighter one.

Without imposing any symmetries, both Higgs doublets can couple to fermions.
The relevant Yukawa interactions in the lepton sector are
\begin{align}
-\mathcal{L}_Y 
&=\bar{l}_{iL}(Y_{1}\Phi_1+Y_{2}\Phi_2 )_{ij} e_{jR}+{\rm h.c.} \nonumber \\
&\ni \bar{e}_{iL}
\left[\frac{y_i}{\sqrt{2}}\delta_{ij}s_{\beta-\alpha}+\frac{1}{\sqrt{2}}\rho_{ij}c_{\beta-\alpha} 
\right] e_{jR}h \nonumber \\
&\quad +\bar{e}_{iL}
\left[\frac{y_i}{\sqrt{2}}\delta_{ij}c_{\beta-\alpha}-\frac{1}{\sqrt{2}}\rho_{ij}s_{\beta-\alpha} 
\right] e_{jR}H \nonumber\\
&\quad+\frac{i}{\sqrt{2}}\bar{e}_{iL}\rho_{ij}e_{jR}A+{\rm h.c.} ~,
\end{align}
where $i,j$ are flavor indices, $Y_{1,2}$ are general 3-by-3 complex matrices, and
\begin{align}
\rho_{ij} = -t_\beta y_i\delta_{ij}+\frac{1}{c_\beta}
\left( V_L^{e\dagger}Y_2V_R^e \right)_{ij} ~,\label{rhoij}
\end{align}
with $V_{L,R}^e$ defined as the unitary matrices that diagonalize the charged lepton Yukawa couplings $Y^{\rm SM}=(Y_1c_\beta+Y_2s_\beta)$, {\it i.e.}, $V_L^{e\dagger}Y^{\rm SM}V_R^e
=Y_D={\rm diag}(y_e, y_\mu, y_\tau)$.
After this diagonalization, the mass terms of the charged leptons are given by $m_i=y_iv/\sqrt{2}$ with $i=e,\mu,\tau$.

Non-diagonal elements of $\rho_{ij}$ are the sources of tree-level FCNH processes.
In the literature, the so-called Cheng-Sher ansatz~\cite{Cheng:1987rs} for $\rho_{ij}$ is adopted
in order to avoid experimental constraints. In our analysis, however, we will not assume it
while obtaining a parameter space that can accommodate both $h\to \mu\tau$ and $g-2$ anomalies.
In general, $Y_{1,2}$ cannot be uniquely determined even though $Y_D$ is known.
In our analysis, we make some working assumption on $Y_{1,2}$ for baryogenesis,
and determine the structure of FCNH couplings $\rho$ at $T=0$.
For later use, we define $\rho_{ij}=|\rho_{ij}|e^{i\phi_{ij}}$.

\section{$h\to \mu \tau$, $(g-2)_\mu$ and EDM}\label{sec:LFV}

In this section, we outline some important consequences of Ref.~\cite{mutau_2HDM_OST} 
to make this paper self-contained.

The Higgs decay to $\mu$ and $\tau$ in the current model occurs at tree level through
$\mu$-$\tau$ interactions (for earlier studies, see Refs.~\cite{LFVHdecay}), and its branching ratio is given by
\begin{align}
{\rm Br}(h\to \mu \tau) 
= \frac{m_h \left(|\rho_{\mu\tau}|^2+|\rho_{\tau\mu}|^2 \right)c_{\beta-\alpha}^2}{16\pi \Gamma_h} ~,
\end{align}
where $m_h = 125$~GeV and $\Gamma_h=4.1$~MeV. It is easy to accommodate
${\rm Br}(h\to \mu \tau)\simeq 0.84\%$ by taking $|\rho_{\mu\tau}| \sim |\rho_{\tau\mu}| \sim \mathcal{O}(0.1)$
and $c_{\beta-\alpha} \sim \mathcal{O}(0.01)$.
It had been shown that such parameter choices did not violate the current experimental bounds on other LFV processes such as $\tau\to\mu\gamma$, $\tau\to \mu\nu\bar{\nu}$, etc~\cite{mutau_2HDM_OST}.
Note that it is sufficient for either $\rho_{\mu\tau}$ or $\rho_{\tau\mu}$ to be nonzero
to explain the $h\to\mu\tau$ excess only.
If both couplings coexist, on the other hand, we can have a one-loop diagram 
that induces $g-2$ and electric dipole moment (EDM) of the muon, denoted by $\delta a_\mu$ and $d_\mu$, respectively. 
Contributions of the neutral Higgs bosons to $\delta a_\mu$ and $d_\mu$ are given by
\begin{align}
\delta a_\mu &= \frac{m_\mu m_\tau{\rm Re}(\rho_{\mu\tau}\rho_{\tau\mu})}{16\pi^2} \nonumber\\
&\quad \times
\left[
	\frac{c_{\beta-\alpha}^2f(r_{h})}{m_h^2}
	+\frac{s_{\beta-\alpha}^2f(r_{H})}{m_H^2}
	-\frac{f(r_{A})}{m_A^2}	
\right] ~, \label{gm2}
\\
\frac{d_\mu}{|e|} &= -\frac{1}{2m_\mu}{\rm Arg}(\rho_{\mu\tau}\rho_{\tau\mu})\delta a_\mu ~,\label{edm}
\end{align}
where
\begin{align}
f(r_{\phi}) &= \frac{-1}{2(1-r_\phi)^2}\left[\frac{2\ln r_\phi}{1-r_\phi}+3-r_\phi\right]
\nonumber \\
&\mathop{\simeq}_{r_\phi\ll1} \ln r_{\phi}^{-1}-\frac{3}{2} ~,
\end{align}
with $r_{\phi}=m_{\tau}^2/m_\phi^2$. 
Non-degeneracy in the neutral Higgs masses and a proper choice of the overall sign 
are essential for obtaining a sufficient $\delta a_\mu$.
In Ref.~\cite{Hagiwara:2011af}, the discrepancy of $(g-2)_{\mu}$ 
between experiment and the SM prediction was estimated as
\begin{align}
\delta a_\mu = a_\mu^{\rm EXP}-a_\mu^{\rm SM} = (26.1\pm 8.0) \times 10^{-10} ~.\label{del_gm2}
\end{align}
As demonstrated in Ref.~\cite{mutau_2HDM_OST}, this deviation could be accommodated 
for $m_{H,A,H^\pm} \in [200, 500]$~GeV
with an appropriate mass hierarchy.
From the EWBG point of view, 
this mass range of the heavy Higgs bosons 
has the right scale for realizing the first-order EWPT, 
though the LFV interactions by themselves do not play a central role in the realization,
as will be shown in the following sections.

The current limit on $d_\mu$ is~\cite{Bennett:2008dy}
\begin{align}
|d_\mu| < 1.9\times 10^{-19} ~e\;{\rm cm} ~,
\end{align}
which is about three orders of magnitude larger than $d_\mu$ estimated with 
${\rm Arg}(\rho_{\mu\tau}\rho_{\tau\mu})=1$ and $\delta a_\mu=3.0\times 10^{-9}$ in Eq.~(\ref{edm}).
In what follows, we will clarify the relationship between CP violation appearing in $d_\mu$ 
and that relevant to BAU.

\section{Baryon number preservation}\label{sec:BP}
The baryon number is generated via the sphaleron process outside the bubble (symmetric phase),
and it can survive if the sphaleron process is sufficiently quenched inside the bubble (broken phase).
To this end, the sphaleron rate in the broken phase, denoted as $\Gamma_B^{(b)}(T)$,
must be smaller than the Hubble constant, $H(T)$.  More explicitly,
\begin{align}
\Gamma_B^{(b)}(T)&\simeq ({\rm prefactor})e^{-E_{\rm sph}(T)/T} \nonumber\\
&< H(T)\simeq 1.66\sqrt{g_*(T)}T^2/m_{\rm P} ~, \label{GamvsH}
\end{align}
where $E_{\rm sph}$ stands for the sphaleron energy, 
$g_*$ is the number of relativistic degrees of freedom in the plasma
($g_*=110.75$ in the 2HDM), and $m_{\rm P}=1.22\times 10^{19}$~GeV.
We parametrize the sphaleron energy as $E_{\rm sph}(T)=4\pi v(T)\mathcal{E}(T)/g_2$
with $g_2$ being the SU(2)$_L$ gauge coupling constant. 
Eq.~(\ref{GamvsH}) is then rewritten as
\begin{align}
\frac{v(T)}{T} > \frac{g_2}{4\pi \mathcal{E}(T)}
\Big[
	42.97+\mbox{log terms}
\Big]\equiv \zeta_{\rm sph}(T) ~.
\label{sph_dec}
\end{align}
One can see that $\zeta_{\rm sph}(T)$ is mostly controlled by $\mathcal{E}(T)$.
The logarithmic corrections in the bracket come from the prefactor of $\Gamma_B^{(b)}$.
To our best knowledge, an extensive study on the prefactor in the 2HDM is still missing.
In the minimal supersymmetric SM case, on the other hand, 
the zero mode factors of the fluctuations about the sphaleron typically amount to about 10\%~\cite{Funakubo:2009eg}.  This is subdominant and, therefore, we will neglect them
in our numerical analysis for simplicity.

We impose Eq.~(\ref{sph_dec}) at a critical temperature $T_C$ at which the effective potential 
has two degenerate minima.
In our analysis, $T_C$ and $v_C\equiv v(T_C)$ are determined using finite-temperature one-loop effective potential with thermal resummation. 
As mentioned in Sec.~\ref{sec:LFV}, $s_{\beta-\alpha}$ is close to 1 in the region of interest to us.
In such a case, we may simplify the analysis of EWPT to a one-dimensional problem with a
single order parameter, as discussed in Refs.~\cite{EWPT_2HDM_1,Kanemura:2004ch}.

As is well-known, the extra heavy Higgs bosons can play a major role in enhancing $v_C/T_C$ in the 2HDM.
In this case, $M^2 \equiv m_3^2/(s_\beta c_\beta)$ must not exceed certain values, depending
on the magnitude of quartic couplings; otherwise, the so-called non-decoupling effects would 
diminish, rendering a suppressed $v_C/T_C$.
Phenomenological consequences of the non-decoupling effects at $T=0$ include 
significant deviations 
in the $h\to\gamma\gamma$ decay width~\cite{Ginzburg:2002wt} 
and the triple Higgs coupling~\cite{2HDM_KOS}.

For the evaluation of $\mathcal{E}$, 
we solve the equations of motion for the sphaleron 
with appropriate boundary conditions~\cite{Manton:1983nd,Klinkhamer:1984di}.
Here, we use the tree-level Higgs potential for simplicity.
In this case, $\zeta_{\rm sph}$ may be underestimated by $\mathcal{O}(10)\%$ since $\mathcal{E}(0)>\mathcal{E}(T_C)$. 
As will be shown below, this approximation does not affect our conclusion.
A detailed analysis of $\zeta_{\rm sph}(T)$ will be given elsewhere.

We note in passing that recent studies show that $\zeta_{\rm sph}(T_C)=1.1 - 1.2$ in the real singlet-extended SM~\cite{Fuyuto:2014yia} and $1.23$ in the scale-invariant 2HDM~\cite{Fuyuto:2015vna} for typical parameter sets.  

\begin{figure}[t]
\center
\includegraphics[width=7cm]{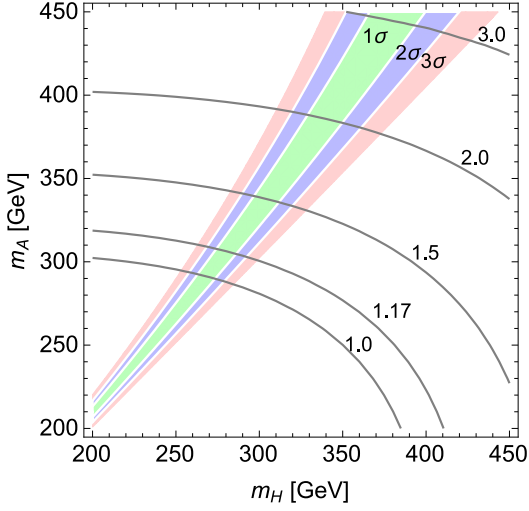}
\caption{Contours of $v_C/T_C$ and $\delta a_\mu$ on the $m_H$-$m_A$ plane.  
We take $m_A=m_{H^\pm}$, $M=100~{\rm GeV}$, $\tan\beta=1$, $c_{\beta-\alpha}=0.006$, $|\rho_{\tau\mu}|=|\rho_{\mu\tau}|$, and $\phi_{\tau\mu}+\phi_{\mu\tau}=\pi/4$.
The solid gray curves are contours of $v_C/T_C = 1.0$, $1.17$ ($=\zeta_{\rm sph}(T_C)$), $1.5$, $2.0$, $2.5$, and $3.0$ from bottom to top.  Areas colored in green, blue and pink correspond to the 1$\sigma$, 2$\sigma$ and 3$\sigma$ regions of $\delta a_\mu$, respectively.}
\label{fig:EWPT}
\end{figure}

In Fig.~\ref{fig:EWPT}, $v_C/T_C$ and $\delta a_\mu$ are shown 
in the ($m_H, m_A$) plane.
We take $m_A=m_{H^\pm}$ to avoid the electroweak $\rho$ parameter constraint,
and choose $c_{\beta-\alpha}=0.006$, $|\rho_{\tau\mu}|=|\rho_{\mu\tau}|$, and
$\phi_{\tau\mu}+\phi_{\mu\tau}=\pi/4$, as favored by the solution of $(g-2)_\mu$ anomaly.
For the remaining parameters, we set $M=100~{\rm GeV}$, $\tan\beta=1$
and $\lambda_6=\lambda_7=0$ as an example.
Contours of $v_C/T_C$ are drawn with the solid gray curves with values of $1.0$, $1.17$, $1.5$, $2.0$, $2.5$, and $3.0$ from bottom to top, where $1.17$ corresponds to $\zeta_{\rm sph}(T_C)$.
Allowed 1$\sigma$, 2$\sigma$ and 3$\sigma$ regions of $\delta a_\mu$ 
are shown by the areas colored in green, blue and pink, respectively.
One can see that the regions satisfying the baryon number preservation condition 
($v_C/T_C>\zeta_{\rm sph}$) and favored by $(g-2)_\mu$ 
have an overlap if $m_A>m_H$.
We note that the $(g-2)_\mu$-favored parameter space can change to the $m_A<m_H$ region
if ${\rm Re}(\rho_{\tau\mu}\rho_{\mu\tau})<0$, as inferred from Eq.~(\ref{gm2}).
However, the allowed region shrinks in this case.

\section{Baryon number generation}\label{sec:BAU}
We closely follow the method given in Refs.~\cite{Huet:1995sh,CTP,Lee:2004we,Chung:2009cb} (see also Refs.~\cite{Cheung:2012pg,Senaha:2013kda,Fuyuto:2015ida}) in estimating the BAU.
In our scenario, the CP-violating source term is induced by interactions between chiral fermions and spacetime-dependent VEV's $v_i(X)$.
As found in Ref.~\cite{Weldon:1999th}, the treatment of the chiral fermion should be taken with some care since its dispersion relation can be significantly modified by thermal plasma.
As a result, the CPV source term for the left-handed (LH) fermion of flavor $i$
induced by the right-handed (RH) fermion of flavor $j$ takes the form 
 \footnote{Note that our result is different from the one shown in Ref. \cite{Blum:2010by}.}
\begin{align}
\lefteqn{S_{i_Lj_R}(X)}\nonumber \\
&= 2C(X)\int_0^\infty\frac{dk\; k^2}{2\pi^2}\;
{\rm Im}
\bigg[
	\frac{(1-n_{jR}^p-n_{iL}^p)Z_{jR}^pZ_{iL}^p}{(\mathcal{E}_{jR}^p+\mathcal{E}_{iL}^p)^2} \nonumber\\
&\quad	+\frac{n_{iL}^{h*}Z_{jR}^pZ_{iL}^{h*}}{(\mathcal{E}_{jR}^p-\mathcal{E}_{iL}^{h*})^2}
	+\frac{n_{jR}^{p*}Z_{jR}^{p*}Z_{iL}^{h}}{(\mathcal{E}_{jR}^{p*}-\mathcal{E}_{iL}^{h})^2}
	+(p \leftrightarrow h)
\bigg] ~,\label{Scpv}
\end{align}
where $C(X)=|Y_{1ij}||Y_{2ij}|\sin\theta_{ij}v^2(X)\partial_{t_X}\beta(X)$ 
with $\theta_{ij}={\rm Arg}(Y_{1ij})-{\rm Arg}(Y_{2ij})$,
$p$ denotes the `particle' mode and $h$ the `hole' mode whose dispersion relations are given by
$\mathcal{E}^{p,h}(k) = E^{p,h}(k)-i\Gamma^{p,h}(k)$, and $Z_{iL,R}^p$ denote the residues.
\footnote{$\mathcal{E}^{p,h}(k)$ can be expressed
in terms of the Lermbert $W$ functions (for details, see Ref.~\cite{Kiessig:2010pr}).}

Here, we remark on the treatment of $\Gamma^{p,h}$. 
In our numerical calculation, $\Gamma^{p,h}$ are approximated as
(constant)$\times T$, as is often done in the calculation of the source terms for scalars and Dirac fermions.
In the chiral fermion case, however, the term without statistical factor in the first term 
of Eq.~(\ref{Scpv}) could cause a logarithmic divergence if $\Gamma^p$ is fixed to a constant 
in the whole range of the $k$ integration, which invalidates the calculation.
Since the correct behavior of $\Gamma^p$ in the large $k$ region may take the form of $T^3 \ln (k/T)/k^2$~\cite{Boyanovsky:2005hk}, 
the integration would be convergent. 
As a simple prescription, we take such a $k$-dependent $\Gamma^h$ for large $k$.
Ambiguities coming from this prescription will be discussed below.

One can also find that $S_{j_Ri_L}=-S_{i_Lj_R}$.
In our study, we consider the case where the 32 and 33 elements of $Y_{1,2}$ play a dominant role in generating the BAU. In this case, only $S_{\tau_L\mu_R}(=-S_{\mu_R\tau_L})$ is relevant.
As mentioned in Sec.~\ref{sec:model}, $Y_{1,2}$ are connected with the $T=0$ observables
through 
$Y_D$ and Eq.~(\ref{rhoij}).
For illustration, we assume that the 12, 13, 21 and 31 elements of $Y^{\rm SM}$ are all zero.
Also, we focus on the case where only $\rho_{\mu\tau}$, $\rho_{\tau\mu}$ 
and $\rho_{\tau\tau}$ take nonzero values among $\rho_{ij}$.

Let us denote the relevant particle number densities as
$Q_3 = n_{t_L}+n_{b_L}$, $T=n_{t_R}$, $B=n_{b_R}$, 
$L _3= n_{\nu_{\tau_L}}+n_{\tau_L}$, $R_3=n_{\tau_R}$, $R_2=n_{\mu_R}$, and
$H=n_{H^+_1}+n_{H^0_1}+n_{H^+_2}+n_{H^0_2}$.
The number density expanded to the leading order in the chemical potential $\mu$
is reduced to $n_{b,f}=T^2\mu k_{b,f}/6$, with $b~(f)$ denoting bosons (fermions).
In the massless case, one finds $k_b=2$ and $k_f=1$.
For massive cases, more precise form of $k_{b,f}$~\cite{Lee:2004we} should be used.

The set of Boltzmann equations may be cast into:
\begin{align}
\begin{split}
\partial_\mu j^\mu_{Q_3} &= -\Gamma_{Y_t}(\xi_{Q_3}+\xi_H-\xi_T)
+\Gamma_{M_t}(\xi_T-\xi_{Q_3})-2\Gamma_{ss}N_5 ~, \\
\partial_\mu j^\mu_T &= \Gamma_{Y_t}(\xi_{Q_3}+\xi_H-\xi_T)
-\Gamma_{M_t}(\xi_T-\xi_{Q_3})+\Gamma_{ss}N_5 ~, \\
\partial_\mu j^\mu_H &= -\Gamma_{Y_t}(\xi_{Q_3}+\xi_H-\xi_T)
+\Gamma_{Y_\tau}(\xi_{L_3}-\xi_H-\xi_{R_3}) \\
&\quad+\Gamma_{Y_{\tau\mu}}(\xi_{L_3}-\xi_H-\xi_{R_2})-\Gamma_H\xi_H ~, \\
\partial_\mu j^\mu_{L_3} &= -\Gamma_{Y_\tau}(\xi_{L_3}-\xi_H-\xi_{R_3})
+\Gamma_{M_\tau}(\xi_{R_3}-\xi_{L_3}) \\
&\quad+\Gamma_{M_{\tau\mu}}^+(\xi_{R_2}+\xi_{L_3})+\Gamma_{M_{\tau\mu}}^-(\xi_{R_2}-\xi_{L_3})\\
&\quad- \Gamma_{Y_{\tau\mu}}(\xi_{L_3}-\xi_H-\xi_{R_2})+S_{\tau_{L}\tau_{R}}+S_{\tau_{L}\mu_{R}} ~,\\
\partial_\mu j^\mu_{R_3} &= \Gamma_{Y_\tau}(\xi_{L_3}-\xi_H-\xi_{R_3})
-\Gamma_{M_\tau}(\xi_{R_3}-\xi_{L_3})-S_{\tau_{L}\tau_{R}} ~, \\
\partial_\mu j^\mu_{R_2} &= \Gamma_{Y_{\tau\mu}}(\xi_{L_3}-\xi_H-\xi_{R_2}) \\
&\quad
-\Gamma_{M_{\tau\mu}}^+(\xi_{R_2}+\xi_{L_3})-\Gamma_{M_{\tau\mu}}^-(\xi_{R_2}-\xi_{L_3})
-S_{\tau_{L}\mu_{R}} ~,
\end{split}
\end{align}
where $\xi_i=n_i/k_i$, $N_5 = 2\xi_{Q_3}-\xi_T+9(Q_3+T)/k_B$, and
$\partial_\mu j^\mu_i=\dot{n}_i-D_i\nabla^2 n_i$ with $D_i$ being a diffusion constant. 
In solving these coupled equations, we utilize the chemical equilibrium conditions 
in light of $\Gamma_{Y_t}^{-1}, \Gamma_{Y_\tau}^{-1}, \Gamma_{Y_{\tau\mu}}^{-1}< \tau_{\rm diff}$, the typical diffusion time for the particles under consideration.
In this case, the above coupled Boltzmann equations are reduced to a single differential equation with 
respect to $H$, as is the case in Ref.~\cite{Huet:1995sh}.

The total LH number density ($n_L$) is 
\begin{align}
n_L&=5Q_3+4T+L_3  \nonumber\\
&=
-\frac{9k_{Q_3}k_T-5k_{Q_3}k_B-8k_Tk_B}{k_H(9k_{Q_3}+9k_T+k_B)}  \nonumber\\
&\quad
+\frac{k_{L_3}D_R(k_{R_3}+k_{R_2})}{k_H\big(D_{L}k_{L_3}+D_R(k_{R_3}+k_{R_2})\big)} ~,
\end{align}
where $D_{L}$ and $D_R$ denote the diffusion constants of the third generation LH lepton doublet
and the RH muon, respectively. 
As pointed out in Ref.~\cite{Chung:2009cb}, the lepton transport is much more efficient 
since it does not suffer from the strong sphaleron washout effect, 
which would make the first term highly suppressed.

With the above $n_L$, the BAU can be estimated as
\begin{align}
n_B = \frac{-3\Gamma_B^{(s)}}{2D_q\lambda_+}
\int_{-\infty}^0dz'~n_L(z')e^{-\lambda_-z'} ~,
\end{align}
where $\lambda_\pm = \big[v_w\pm\sqrt{v_w^2+4\mathcal{R}D_q}\big]/2D_q$,
$v_w$ represents the expanding speed of the bubble wall, $D_q$ is the diffusion constant of the quarks, 
$\Gamma_B^{(s)}$ is the sphaleron rate
in the unbroken phase, and $\mathcal{R}=15\Gamma_B^{(s)}/4$.
In the following, we will characterize the baryon asymmetry by $Y_B \equiv n_B / s$, where $s = (2\pi^2/45)g_*$ is the entropy density at temperature $T$, 
and quote $Y_B^{\rm obs}= 8.59\times 10^{-11}$, the central value given by the Planck Collaboration~\cite{Ade:2013zuv}.

In the estimate of BAU, we employ the formulas given in Ref.~\cite{DiffConst} 
for the diffusion constants and thermal widths of LH and RH leptons, and assume $D_h=D_{L}$. The particular values that we use here are summarized in Table. \ref{tab:input}.

\begin{table}[t]
\center
\begin{tabular}{cccc}
\hline\hline
$T_C=99.2$ GeV & $v_C=214.9$ GeV & $v_w=0.1$ & $\Delta\beta=0.05$  \\
$m_{\tau_L}=0.21T$ & $m_{\mu_R}=0.12T$ & $\Gamma_{\tau_L}=0.034T$ & $\Gamma_{\mu_R}=0.015T$ \\
$D_q=8.9/T$  & $D_h=101.9/T$ & $D_L=101.9/T$ & $D_R=377.1/T$ \\ 
\hline\hline
\end{tabular}
\caption{Input parameters for the BAU estimate.}
\label{tab:input}
\end{table}


\begin{figure}[t]
\center
\includegraphics[width=7cm]{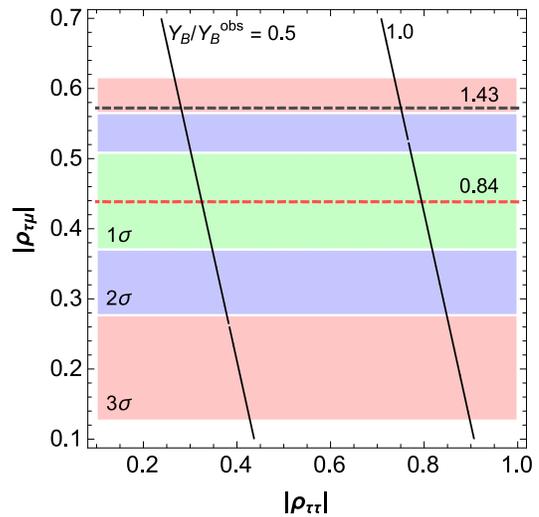}
\caption{Contours of $Y_B/Y_B^{\rm obs}$, ${\rm Br}(h\to \mu\tau)$ and $\delta a_{\mu}$ 
in the ($|\rho_{\tau\tau}|$, $|\rho_{\tau\mu}|$) plane. We set $m_H=350$ GeV, 
$m_A=m_{H^\pm}=400$ GeV, $M=100$ GeV, $c_{\beta-\alpha}=0.006$, 
$|\rho_{\tau\mu}|=|\rho_{\mu\tau}|$,
$\phi_{\tau\mu}=-5\pi/4$, $\phi_{\mu\tau}=\pi/4-\phi_{\tau\mu}$ and $\phi_{\tau\tau}=\pi/2$.}
\label{fig:BAU}
\end{figure}

In Fig.~\ref{fig:BAU}, we show $Y_B/Y_B^{\rm obs}$, ${\rm Br}(h\to \mu\tau)$ and $\delta a_{\mu}$ as functions of $|\rho_{\tau\tau}|$ and $|\rho_{\tau\mu}|$. 
Here we take the same input parameters as those in Fig.~\ref{fig:EWPT} 
and the heavy Higgs boson masses $m_H=350~$GeV and $m_A=m_{H^{\pm}}=400~$GeV, leading to $v_C/T_C=214.9~{\rm GeV}/99.2~{\rm GeV}=2.17$. 
As an ansatz for $Y^{\rm SM}$, we consider
\begin{align}
Y^{\rm SM}  =
\begin{pmatrix}
\sqrt{2}m_e/v & 0 & 0 \\
0 & 3.31\times10^{-3} & -6.81i \times10^{-4}  \\
0 & 8.91i\times10^{-3} & 3.70\times10^{-3}
\end{pmatrix},
\end{align}
which is diagonalized
by
\begin{align}
V_R^e&=
\begin{pmatrix}
1 & 0 & 0\\
0 & -0.365i & -0.931 i \\
0 & -0.931 & 0.365
\end{pmatrix},\\
V_L^e&=
\begin{pmatrix}
1 & 0 & 0\\
0 & -0.945i & -0.327i \\
0 & -0.327 & 0.945
\end{pmatrix}.
\end{align}
Also, we take $\phi_{\tau\tau}=\pi/2$.
Note that each of $Y_{1,2}$ is fixed by $V_L^{e\dagger}Y^{\rm SM}V_R^e =Y_D$ and Eq.~(\ref{rhoij})
once $\rho_{ij}$ are given. 

The solid lines represent the contours of $Y_B/Y_B^{\rm obs}=0.5$ (left) and $1.0$ (right).
The dashed lines in black and red represent ${\rm Br}(h\to \mu\tau)=1.43\%$ and $0.84\%$, respectively. The colored regions with the same color scheme as in Fig.~\ref{fig:EWPT}
explain the $(g-2)_{\mu}$ anomaly. 
In our setup, the dominant contribution to the CP-violating source term comes from 
$S_{\tau_L\mu_R}$ which is induced by the 32 elements of $Y_1$ and $Y_2$.
We find that these off-diagonal elements have stronger dependences on $|\rho_{\tau\tau}|$ 
than $|\rho_{\tau\mu}|$ and $|\rho_{\mu\tau}|$.
Under rather generous assumptions for the bubble wall profile, 
the generated BAU can reach its observed value  
for $|\rho_{\tau\mu}|\simeq 0.1-0.6$ and $|\rho_{\tau\tau}|\simeq 0.8-0.9$.
Our main conclusion is that there is a parameter space that is consistent with 
both experimental anomalies of $h\to\mu\tau$ and $(g-2)_\mu$ as well as the observed BAU.

Some comments on the theoretical uncertainties are in order.
(1) We take the same size of $|\Delta \beta|$ as in Ref.~\cite{Liu:2011jh} as a reference value. 
However, quantitative studies of it in the 2HDM are still absent. 
In the MSSM, $\Delta \beta=\mathcal{O}(10^{-2}-10^{-4})$ depending on $m_A$~\cite{Moreno:1998bq}.
In the next-to-MSSM, $\Delta \beta$ can reach $\mathcal{O}(0.1)$ in some parameter space~\cite{Kozaczuk:2014kva}.
We should note that $Y_B$ is mostly subject to the uncertainties of $\Delta \beta$ among others
since it is linearly proportional to $\Delta \beta$.
(2) Studies on $v_w$ can be found in Refs.~\cite{Megevand:2009gh,Ahmadvand:2013sna,Konstandin:2014zta,Kozaczuk:2015owa}, which suggest that $0.1\lesssim v_w \lesssim 0.6$ in non-SUSY models. 
For stronger first-order EWPT, $v_w$ tends to be large and may reduce $Y_B$ to some extent. 
In our analysis, we simply adopt the lowest value as the most generous choice.
A more precise determination of $v_w$ using our input parameters is definitely indispensable to 
obtaining more precise $Y_B$.
(3) As mentioned above, the treatment of $\Gamma^h$ is somewhat tricky. 
It is found that $Y_B$ may change by a factor of a few or more, depending on 
at which scale the $k$-dependent $\Gamma^h$ is put in.
(4) The VEV-insertion method used here is vulnerable to theoretical uncertainties
(see, {\it e.g.}, Refs.~\cite{Morrissey:2012db,Inoue:2015pza})
and may lead to an overestimated BAU compared to an all-order VEV resummation method~\cite{Carena:2000id,Carena:2002ss,BAU_woVEVinsertion}.
However, a satisfactory formalism of the $Y_B$ calculation beyond that is still not available,
and hence the error associated with the approximation cannot be properly quantified. 

In summary, the observed $Y_B$ can be marginally produced with the generous choice 
of the input parameters.
However, a definitive statement cannot be made until the above-mentioned various theoretical uncertainties are fully under control. 

\begin{figure}[t]
\center
\includegraphics[width=7cm]{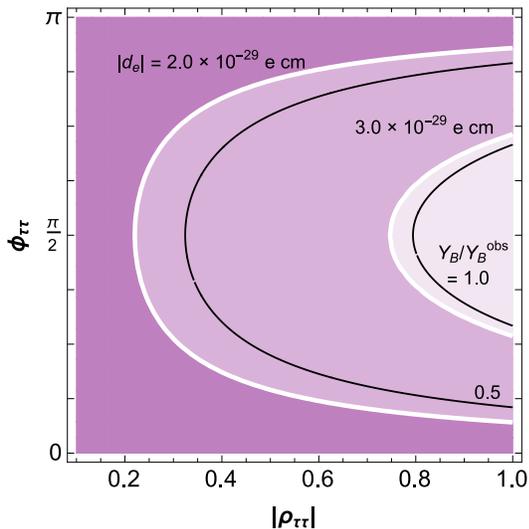}
\caption{Correlations between BAU and electron EDM 
in the ($|\rho_{\tau\tau}|$, $\phi_{\tau\tau}$) plane.
The contours shown in white are $|d_e|=2.0\times 10^{-29}~e~{\rm cm}$ (left)
and $3.0\times 10^{-29}~e~{\rm cm}$ (right).
On the other hand, $Y_B/Y_B^{\rm obs}=0.5$ (left) and 1.0 (right)
are denoted by the contours in black.}
\label{fig:BAUvsEDM}
\end{figure}

We now turn to another experimental constraint.
As $|\rho_{\tau\tau}|$ increases, ${\rm Br}(\tau \to \mu \gamma)$ gets enhanced 
and, for $|\rho_{\tau\tau}|\gtrsim 0.1$, exceeds the current experimental upper bound, 
${\rm Br}(\tau \to \mu \gamma)<4.4\times 10^{-8}$~\cite{TauMuGamEXP}.
However, as demonstrated in Ref.~\cite{mutau_2HDM_OST}, 
an accidental cancellation between the one-loop and two-loop contributions could occur in ${\rm Br}(\tau \to \mu \gamma)$ 
if the new top Yukawa coupling (denoted as $\rho_{tt}$) took a nonzero value.
In the current case, $|\rho_{tt}|\simeq 0.5$ with $\phi_{tt}\simeq \phi_{\tau\tau}$
gives ${\rm Br}(\tau\to\mu\gamma)\simeq2\times10^{-8}$
for $|\rho_{\tau\tau}|\simeq1$.
The impact of $\phi_{tt}$ on $Y_B$ highly depends on the Yukawa structure 
of the top quark.  With our assumption, the top quark does not provide dominant CP violation for the baryogenesis.

Since $\rho_{\tau\tau}$ and $\rho_{tt}$ are complex, they may induce an electron EDM
through two-loop Barr-Zee diagrams, among which one Higgs-photon mediated loop diagram is dominant.
In Fig.~\ref{fig:BAUvsEDM}, $|d_e|$ and $Y_B/Y_B^{\rm obs}$ are plotted
as functions of $|\rho_{\tau\tau}|$ and $\phi_{\tau\tau}$. The contours in white represent $|d_e|=2.0\times 10^{-29}~e~{\rm cm}$ (left) and $3.0\times 10^{-29}~e~{\rm cm}$ (right). 
$|d_e|$ with $|\rho_{\tau\tau}|=1$ and $\phi_{\tau\tau}=\pi/2$ reaches a maximal value
of $3.5\times10^{-29}~e~{\rm cm}$
which is slightly smaller than the current bound on electron EDM,
$|d_e|<8.7\times 10^{-29}~e~{\rm cm}$~\cite{Baron:2013eja}.
This is due to the facts that the heavy Higgs boson couplings to the electron is suppressed 
by $c_{\beta-\alpha}$ and that the extra Yukawa coupling $\rho_{ee}$ is absent. 
It should be noted that the complex phases of $\rho_{\tau\mu}$ and $\rho_{\mu\tau}$ do not contribute to $d_e$
via the Higgs-photon mediated Barr-Zee diagrams 
since the internal photon line cannot change the fermion flavors.   
As discussed in Sec.~\ref{sec:LFV}, on the other hand, the muon EDM is induced by those FCNH couplings at one-loop level, giving rise to $|d_\mu|\simeq 3\times 10^{-22}~e~{\rm cm}$
for the current parameter set.

We also find that $\tau \to \mu\nu\bar{\nu}$ can give some constraints on 
$|\rho_{\tau\mu}|$ and $|\rho_{\mu\tau}|$ as well as $m_{H^\pm}$ and can be similar to the constraint coming from ${\rm Br}(h\to\tau\mu)<1.43$.

\section{Conclusion}\label{sec:conclusion}

We have studied electroweak baryogenesis in the general framework of the two-Higgs doublet model in light of the $h\to \mu\tau$ and $(g-2)_{\mu}$ anomalies. 
In this model, the heavy Higgs bosons with the appropriate $\mu$-$\tau$ flavor violation 
can accommodate the above two anomalies. 
At the same time, these extra Higgs bosons can induce a strong first-order electroweak phase transition 
as required for successful electroweak baryogenesis.

It is found that the $\mu$-$\tau$ flavor-violating lepton sector 
has a great potential to generate sufficient baryon asymmetry of the Universe within the theoretical uncertainties.
In this scenario, the interplay between $\rho_{\tau\mu}/\rho_{\mu\tau}$ and $\rho_{\tau\tau}$ 
is crucially important. Our analysis suggests that $Y_B/Y_B^{\rm obs}\simeq 1$
for $|\rho_{\tau\mu}|\simeq 0.1-0.6$ and $|\rho_{\tau\tau}|\simeq 0.8-0.9$ with $\mathcal{O}(1)$ 
CP-violating phases.
To suppress ${\rm Br}(\tau\to\mu\gamma)$, 
a cancellation mechanism has to be at work, which additionally predicts $|\rho_{tt}|\simeq0.5$ 
and $\phi_{tt}\simeq\phi_{\tau\tau}$.
Since future experimental sensitivities of ${\rm Br}(\tau\to\mu\gamma)$ and $|d_e|$
are about $1\times 10^{-9}$~\cite{Aushev:2010bq} and $10^{-30}~e~{\rm cm}$~\cite{future_eEDM}, respectively, our scenario could be fully tested.


\begin{acknowledgments}
E.S. thanks Yoshimasa Hidaka for useful discussions.
This work is supported in part by the Ministry of Science and Technology of R.O.C. under Grant Nos. 104-2628-M-002-014-MY4 (CWC) and 104-2811-M-008-011 (ES), and in part by the Research Fellowships of the Japan Society for the Promotion of Science for Young Scientists, No. 15J01079 (KF).
\end{acknowledgments}


\end{document}